\documentclass[12pt]{article}
\usepackage{epsfig}
\usepackage{rotating}
\usepackage{lscape}
\usepackage{amsmath}
\usepackage[usenames]{color}

\begin{document}

\begin{center}

\vspace*{1.0cm}

 {\Large \bf{First search for $2\varepsilon$ and $\varepsilon\beta^+$
decay of $^{162}$Er and new limit on $2\beta^{-}$ decay of
$^{170}$Er to the first excited level of $^{170}$Yb}}

\vskip 1.0cm

{\bf P.~Belli$^{a}$, R.~Bernabei$^{a,b,}$\footnote{Corresponding
author. {\it E-mail address:} rita.bernabei@roma2.infn.it
(R.~Bernabei).}, R.S.~Boiko$^{c,d}$, F.~Cappella$^{e}$,
V.~Caracciolo$^{f}$, R.~Cerulli$^{a}$, F.A.~Danevich$^{c}$,
A.~Incicchitti$^{e,g}$, B.N.~Kropivyansky$^{c}$,
M.~Laubenstein$^{f}$, S.~Nisi$^{f}$, D.V.~Poda$^{c,h}$,
O.G.~Polischuk$^{c}$, V.I.~Tretyak$^{c}$}

\vskip 0.3cm

$^{a}${\it INFN sezione Roma ``Tor Vergata'', I-00133 Rome, Italy}

$^{b}${\it Dipartimento di Fisica, Universit$\grave{a}$ di Roma
``Tor Vergata'', I-00133 Rome, Italy}

$^{c}${\it Institute for Nuclear Research, 03028 Kyiv, Ukraine}

$^{d}${\it National University of Life and Environmental Sciences
of Ukraine, 03041 Kyiv, Ukraine}

$^{e}${\it INFN sezione Roma, I-00185 Rome, Italy}

$^{f}${\it INFN, Laboratori Nazionali del Gran Sasso, I-67100
Assergi (AQ), Italy}

$^{g}${\it Dipartimento di Fisica, Universit$\grave{a}$ di Roma
``La Sapienza'', I-00185 Rome, Italy}

$^{h}${\it CSNSM, Universit\'e Paris-Sud, CNRS/IN2P3, Universit\'e
Paris-Saclay, 91405 Orsay, France}

\end{center}

\vskip 0.5cm

\begin{abstract}
The first search for double electron capture ($2\varepsilon$) and
electron capture with positron emission ($\varepsilon\beta^+$) of
$^{162}$Er to the ground state and to several excited levels of
$^{162}$Dy was realized with 326 g of highly purified erbium
oxide. The sample was measured over 1934 h by the ultra-low
background HP Ge $\gamma$ spectrometer GeCris (465
cm$^3$) at the Gran Sasso underground laboratory. No effect was
observed, the half-life limits were estimated at the level of
$\lim T_{1/2}\sim 10^{15}-10^{18}$ yr. A possible resonant $0\nu
KL_1$ capture in $^{162}$Er to the $2^+$ 1782.7 keV excited state
of $^{162}$Dy is restricted as $T_{1/2}\geq5.0\times 10^{17}$ yr
at 90\% C.L. A new improved half-life limit $T_{1/2}\geq4.1\times
10^{17}$ yr was set on the $2\beta^-$ decay of $^{170}$Er to the
$2^+$ 84.3 keV first excited state of $^{170}$Yb.

\end{abstract}

\vskip 0.4cm


\vskip 0.4cm


\section{INTRODUCTION}

The neutrinoless double beta ($0\nu2\beta$) decay is forbidden in the
Standard Model of particle physics (SM) since the process violates
the lepton number and allows to investigate if the neutrino is a
Majorana particle. Therefore, searches for this decay are
considered as an unique way to study the properties of the neutrino and of the
weak interaction, to test the lepton number violation, to search for
effects beyond the SM
\cite{Barea:2012,Rodejohann:2012,Deppisch:2012,Bilenky:2015,Delloro:2016,Vergados:2016}.
While the two neutrino ($2\nu$) mode of $2\beta^-$ decay has been
already observed in several nuclei with the half-lives
$T^{2\nu2\beta^-}_{1/2}\sim 10^{18}-10^{24}$ yr
\cite{Tretyak:2002,Saakyan:2013,Barabash:2015}, the $0\nu2\beta^-$
decay is still under investigation. Even the most sensitive
experiments give only half-life limits on the decay at the level
of $\lim T^{0\nu2\beta^-}_{1/2}\sim 10^{24}-10^{26}$ yr (we refer
reader to the reviews
\cite{Delloro:2016,Tretyak:2002,Elliott:2012,Giuliani:2012,Cremonesi:2014,Sarazin:2015}
and the recent original works
\cite{Arnold:2015,Gando:2016,Albert:2018,Alduino:2018,Aalseth:2018,Agostini:2018,Azzolini:2018}).

The achievements in investigations of the double beta plus
processes, such as double electron capture ($2\varepsilon$),
electron capture with positron emission ($\varepsilon\beta^+$) and
double positron decay ($2\beta^+$) are much more modest
\cite{Tretyak:2002,Maalampi:2013,Blaum:2018}. The
``gap'' can be explained by the typically very low isotopic
abundance of the double beta plus isotopes, that does not exceed
1\%, and the suppression of the decay probabilities by small phase
space factors. This leads to a much lower sensitivity of the
experiments to the effective Majorana neutrino mass. Even the
allowed two neutrino double electron capture is not observed
surely. There are only indications on the double electron capture
in $^{130}$Ba \cite{Meshik:2001,Pujol:2009} and $^{78}$Kr
\cite{Gavrilyuk:2013, Ratkevich:2017}.

At the same time, the need to develop experimental methods to
search for double beta plus processes is supported by the
capability to distinguish between two possible
mechanisms of the $0\nu2\beta^{-}$ decay if observed: whether it
is due to the light Majorana neutrino mass or due to the
right-handed currents' admixture in the weak interaction
\cite{Hirsch:1994}. Another argument in favor of the neutrinoless
double electron capture investigations is the possibility of
resonant enhancement of the capture rate due to a mass degeneracy
between the initial and final nucleus
\cite{Winter:1955,Voloshin:1982,Bernabeu:1983}.

The isotope $^{164}$Er was proposed as a candidate to search for
the Majorana neutrino mass in the resonant $0\nu2\varepsilon$
process \cite{Sujkowski:2004,Krivoruchenko:2011,Fang:2012}. The
$Q_{2\beta}$ value of the neutrinoless double electron capture
transitions in $^{164}$Er was precisely measured by Penning-trap
mass-ratio spectrometry as 25.07(12) keV. The value results in a
rather long theoretical prediction for the half-life of
$\sim10^{30}$ yr for a 1 eV effective Majorana neutrino mass
\cite{Eliseev:2011b}. Taking into account that the
sensitivity of the most recent $0\nu2\beta^-$ experiments is
almost one order magnitude higher (e.g., the KamLAND-Zen
experiment already reached an effective Majorana neutrino mass
sensitivity $\lim \langle m_{\nu}\rangle \sim 0.1$ eV
\cite{Gando:2016}) the corresponding theoretical half-life for
$^{164}$Er is on the level of $\sim10^{32}$ yr. Nevertheless,
despite the precision mass measurements indicate that the actually
most promising double electron capture candidates are $^{152}$Gd,
$^{156}$Dy and $^{190}$Pt
\cite{Eliseev:2011b,Eliseev:2011c,Eibach:2016}, $^{164}$Er remains
an interesting nucleus in the list of resonant neutrinoless double
electron capture candidates. So that was the
reason to investigate the radiopurity level of erbium, and to estimate the
possibilities of erbium purification from radioactive elements. In
addition to $^{164}$Er, erbium contains two other potentially
double beta active isotopes: the double beta plus isotope,
$^{162}$Er, and the $2\beta^-$ $^{170}$Er. Characteristics of
these isotopes are given in Table \ref{tab:2bisotopes}.

\nopagebreak
\begin{table}[!ht]
\caption{Characteristics of $2\beta$ isotopes of erbium.}
\begin{center}
\begin{tabular}{|l|l|l|l|}
\hline
 $2\beta$ transition                & $Q_{2\beta}$ (keV)                & Isotopic abundance (\%) \cite{Meija:2016} & Decay channel \\
 \hline
$^{162}$Er$\rightarrow$$^{162}$Dy   & 1846.95(30) \cite{Eliseev:2011a}  &   0.139(5)                                & $2\varepsilon$, $\varepsilon\beta^+$  \\
 \hline
$^{164}$Er$\rightarrow$$^{164}$Dy   & 25.07(12) \cite{Eliseev:2011b}    &   1.601(3)                                & $2\varepsilon$ \\
 \hline
$^{170}$Er$\rightarrow$$^{170}$Yb   & 655.2(15) \cite{Wang:2017}       &   14.910(36)                              & $2\beta^-$  \\

\hline
\end{tabular}
  \label{tab:2bisotopes}
\end{center}
\end{table}

Unfortunately, the low energy release expected in the double
electron capture of $^{164}$Er does not allow the search for the decay
by using the low-background $\gamma$ spectrometry applied in the
present work. Nevertheless, in addition to the radiopurity
investigations of the erbium sample, we have used the data of the low-background
measurements to derive new limits on double beta processes in
$^{162}$Er and $^{170}$Er with emission of 511 keV $\gamma$ quanta
after $\beta^+$ annihilation, or $\gamma$ quanta expected in the
de-excitation of daughter nuclei.

\begin{figure}[htb]
\begin{center}
 \mbox{\epsfig{figure=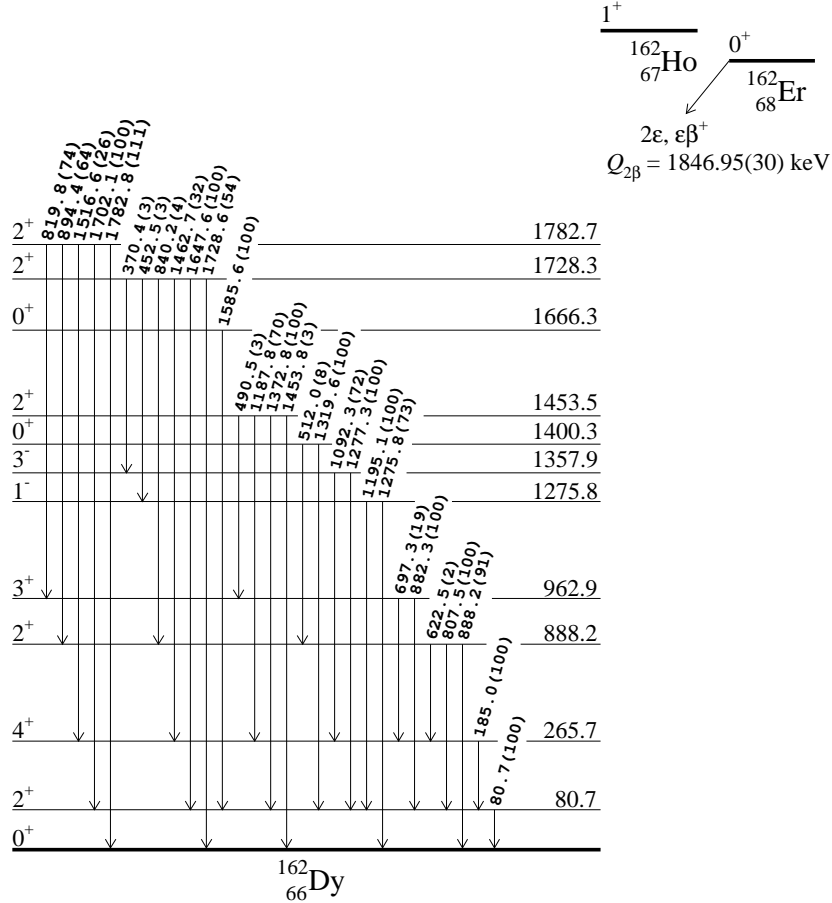,height=12.0cm}}
\vspace{-0.3cm} \caption{Simplified decay scheme of $^{162}$Er
\cite{Reich:2007}. The energies of the excited levels and of the
emitted $\gamma$ quanta are in keV (relative intensities of
$\gamma$ quanta, rounded to percent, are given in parentheses;
only the $\gamma$ transitions with the relative intensity of more
than 2\% are shown).}
\end{center}
\label{fig:162er}
\end{figure}

A simplified scheme of the double beta decay of $^{162}$Er is
presented in Fig. \ref{fig:162er} (in the daughter $^{162}$Dy only the $\gamma$ transitions
with relative intensities of more than 2\%
are shown).

\begin{figure}[htb]
\begin{center}
 \mbox{\epsfig{figure=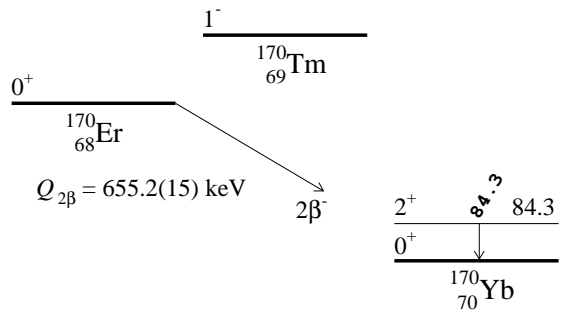,height=4.cm}}
\vspace{-0.3cm} \caption{Simplified decay scheme of $^{170}$Er
\cite{Baglin:2002}.}
\label{fig:170er}
\end{center}
\end{figure}

The double beta decay of $^{170}$Er is possible to the ground and to the
first $2^+$ excited level of $^{170}$Yb with energy 84.3 keV (see
Fig. \ref{fig:170er}).

\clearpage

\section{EXPERIMENT}

\subsection{Purification of erbium oxide}

The erbium oxide (Er$_2$O$_3$) for the experiment was provided by
the Stanford Materials Corporation. The purity level of the
material was $\>99.5\%$ (TREO, total rare earth oxides) and
$\>99.999\%$ (Er$_2$O$_3$/TREO). The compound was examined by
Inductively Coupled Plasma Mass Spectrometer (ICP-MS, model
Element II from Thermo Fisher Scientific, Waltham, Massachusetts,
USA). In order to overcome the drawback related to the well known
isobaric interferences, K and Fe were measured respectively in
High Resolution (HR) and Medium Resolution mode. All the other
elements have been analyzed in Low Resolution (LR) setting (see
Table 2). The initial contamination of the material by radioactive
lanthanide elements is in agreement with the producer
specification (concentrations of La and Lu less than 0.1 ppm).
Thorium and uranium were also in the material as 1.6 ppb and 1.9
ppb, respectively.

\clearpage \nopagebreak
\begin{table}[!h]
\caption{Contamination of the erbium oxide sample measured by
ICP-MS before and after the purification, as well as contamination
of the Er$_2$O$_3$ sediment after the fractional precipitation
stage of the purification (see text). Errors on the measured
values are at the level of 30\% since the ICP-MS measurements have
been carried out in Semi-Quantitative mode.}
\begin{center}
\begin{tabular}{|l|l|l|l|}

 \hline
  Element       & \multicolumn{3}{c|}{Concentration (ppb)} \\
 \cline{2-4}
  ~             & Initial material      &  Sediment after           & After purification \\
  ~             & before purification   &  fractional precipitation & by the liquid-liquid \\
  ~             & ~                     &  ~                        & extraction method \\
  \hline
 K              &  $<2000$              & 424                       & 283 \\
  \hline
 Fe             & -                     & 1190                      & 1797 \\
  \hline
 Pb             & 10000                 &  56                       & 322 \\
  \hline
 La             &  100                  & 16                        & 17 \\
  \hline
 Lu             &  80                   & 144                       & 99 \\
 \hline
 Th             &  1.6                  &  13                       & $<0.3$   \\
 \hline
 U              &  1.9                  &  $<0.2$                   & $<0.2$  \\
 \hline

\end{tabular}
\end{center}
\label{tab:icpms}
\end{table}

The radioactive contamination of a 517 g sample of the erbium
oxide was measured over 340 h with the p-type ultra-low background
high purity germanium (HPGe) $\gamma$ spectrometer
GePV with an active volume of 363 cm$^3$ at the
STELLA facility of the Gran Sasso underground laboratory of the
INFN (Italy). The energy resolution of the detector is FWHM~$=1.8$
keV for 1333 keV $\gamma$ quanta of $^{60}$Co, the relative
efficiency is 91\% \cite{ML_2017}. The detection efficiencies to
$\gamma$ quanta emitted in the decay of the radioactive
contamination nuclides were calculated with the GEANT4 simulation
package \cite{Agos03,Alli06} with initial kinematics given by the
DECAY0 event generator \cite{DECAY0a,DECAY0b}. The results of the
measurements (denoted as "Before purification") are presented in
Table \ref{tab:rcont}. It should be noted that the
low sensitivity of the measurements before purification
(particularly to $^{234}$Th) is due to a much lower detection
efficiency to low energy $\gamma$ quanta and higher background
counting rate of the GePV detector. E.g., presence of $^{234}$Th
in the sample was estimated by searching for $\gamma$ quanta with
energies 92.4 keV and 92.8 keV.

\clearpage \nopagebreak
\begin{table}[!h]
\caption{Radioactive contamination of the erbium oxide sample
before and after the purification, measured with the help of an
ultra-low background HP Ge $\gamma$ spectrometer. The upper limits
are presented at 90\% C.L., the uncertainties are given at
$\approx 68\%$ C.L.}
\begin{center}
\begin{tabular}{|l|l|l|l|}

 \hline
  Chain     & Nuclide       & \multicolumn{2}{c|}{Activity (mBq/kg)} \\
 \cline{3-4}
  ~         & ~             & Before purification   & After purification \\
  \hline
 ~          & $^{40}$K      & $\leq 27$             & $\leq 1.7$ \\
 ~          & $^{137}$Cs    & $\leq 2.1$            & $1.4\pm0.3$\\
 ~          & $^{176}$Lu    & $6\pm1$               & $4.2\pm0.4$\\
\hline
 $^{232}$Th & $^{228}$Ra    & $\leq 7.2$            & $\leq 1.0$ \\
 ~          & $^{228}$Th    & $5\pm 2$              & $\leq 1.1$ \\
 \hline
 $^{235}$U  & $^{235}$U     & $\leq 12$             & $\leq 1.6$\\
\hline
 $^{238}$U  & $^{226}$Ra    & $6\pm 2$              & $1.1\pm0.4$\\
 ~          & $^{234}$Th    & $\leq 1800$           & $\leq 91$\\
 ~          & $^{234m}$Pa   & $\leq 74$             & $\leq 16$\\
\hline

\end{tabular}
\end{center}
\label{tab:rcont}
\end{table}

Traces of lutetium ($^{176}$Lu), radium ($^{226}$Ra) and thorium
($^{228}$Th) were detected in the sample with activities at the
level\footnote{We would like to emphasize that the
results of the ICP-MS and $\gamma$ spectrometry on thorium and
lutetium are in a good agreement (1.6 ppb of Th corresponds to
an activity of $^{228}$Th 6 mBq/kg, assuming the equilibrium of the
$^{232}$Th chain; 80 ppb of Lu corresponds to 4 mBq/kg of
$^{176}$Lu).} of $5-6$ mBq/kg. Therefore, an
additional purification of the material was decided.

The following scheme of purification procedure was applied to the
Er$_2$O$_3$ purification: 1) dissolving of Er$_2$O$_3$; 2)
fractional precipitation of Er(OH)$_3$ sediment; 3) liquid-liquid
extraction; 4) precipitation of Er(OH)$_3$; and 5) final
recovery of Er$_2$O$_3$.

As a first step, diluted nitric acid was added to a suspension
of Er$_2$O$_3$ in deionized water to obtain a homogeneous aqueous
solution of erbium. The initial amounts of water and nitric acid
were calculated to obtain an acidic solution with a concentration of
Er$^{3+}$ at the level of 1.5 mol/L.

The fractional precipitation of erbium from the acidic solution
was used to co-precipitate impurities like Th, taking into account
that hydroxides of thorium are precipitated at a lower pH level
than erbium. Ammonia gas has been injected into the solution till
the pH reached 6.5 that led to the fractional precipitation of
erbium hydroxide. Then, the amorphous Er(OH)$_3$ sediment was
separated from the supernatant liquid using a centrifuge, and was
annealed to Er$_2$O$_3$. The 43 g of erbium oxide were obtained,
which is 8.3\% of the initial mass. The oxide was analyzed by the
ICP-MS to check the efficiency of the co-precipitation of the
impurities (see Table \ref{tab:icpms}, ``Sediment after fractional
precipitation''). The contamination of the sediment testifies the
efficiency of the purification process. For instance, the
concentration of Th increased in the sediment by a factor of
$\approx8$. However, it is a rather minor figure taking into
account the requirements of the double beta experiments aiming at
the achievement of an as low as possible level of background that
is determined by the radioactive contamination of the sample.
Therefore, the liquid-liquid extraction method was applied for
further purification of the material. The sediment
was excluded from the further purification process by the
liquid-liquid extraction method since it accumulated impurities of
the initial material.

The liquid-liquid extraction method \cite{Boiko:2017} proved to be
the most effective one for the purification of lanthanides
solutions from traces of uranium and thorium. To apply
liquid-liquid extraction to the erbium solution, it was acidified
with diluted nitric acid to pH~$=1$. Tri-n-octylphosphine oxide
(TOPO) has been utilized as ``soft'' organic complexing agent for
binding U and Th, while toluene was used as liquor solvent.
Considering the chemically very low concentration of the traces in
the solution, the concentration of TOPO in toluene  did not exceed
0.1 mol/L. The two immiscible liquids (aqueous solution and
organic solution) were placed in a separation funnel in the
volumetric ratio of 1:1 and shaken for a few minutes. The uranium and
thorium interact with TOPO forming organo-metallic complexes that
have much higher solubility in organic phase than in water
solution. This leads to the extraction of U and Th into the
organic liquid. After the separation of the purified aqueous
solution, the erbium was completely precipitated in form of hydroxide
using ammonia. The impurities as alkali and alkali-earth cations
were left in the supernatant liquid. Sediments were separated,
dried and annealed at 900 $^{\circ}$C for a few hours. Finally,
398 g of purified material were obtained, that is 77\% of the
initial material. The purified material was analyzed by the ICP-MS
as reported in Table \ref{tab:icpms}.

\subsection{Low counting experiment}

\begin{figure}[htb]
\begin{center}
 \mbox{\epsfig{figure=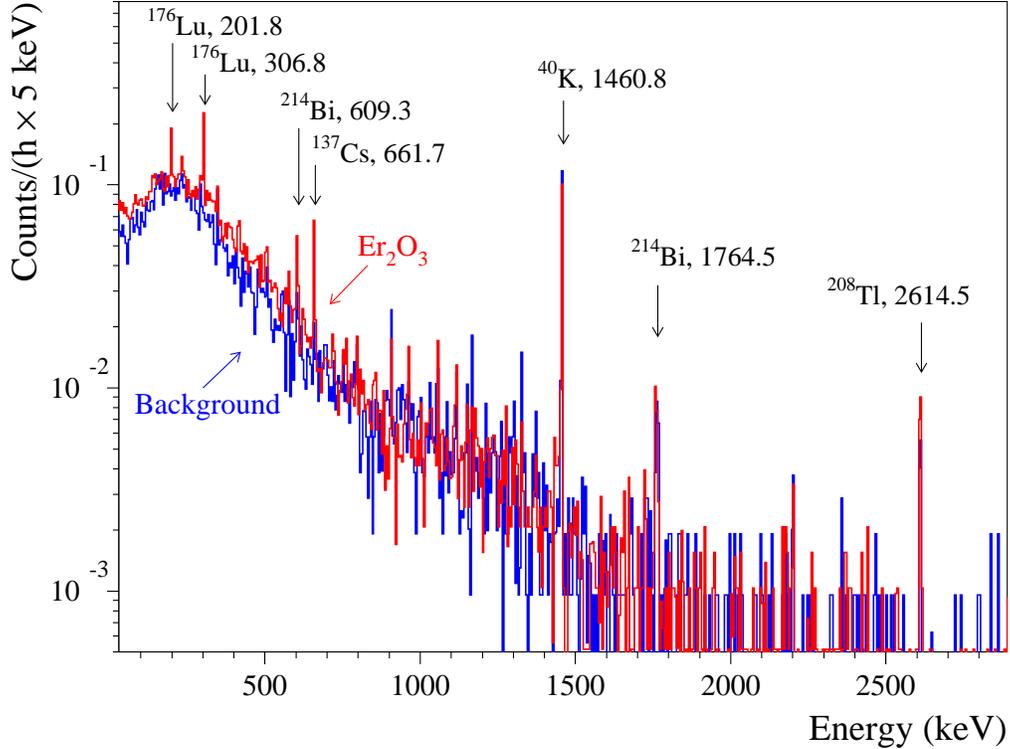,height=10.0cm}}
\vspace{-0.3cm} \caption{(Color online) Energy spectra measured
with the ultra-low background HPGe $\gamma$ spectrometer with the purified
Er$_2$O$_3$ sample over 1934 h (Er$_2$O$_3$) and without sample
over 1046 h (Background). The energies of the $\gamma$ peaks are
in keV.}
 \label{fig:bg}
\end{center}
\end{figure}

The experiment was carried out at the STELLA facility by using the
ultra-low background HPGe detector GeCris with a volume of 465
cm$^3$. The detector is shielded by low radioactive lead
($\approx25$ cm), copper ($\approx 5$ cm), and in the innermost
part by archaeological Roman lead ($\approx 2.5$ cm). The set-up
is enclosed in an air-tight poly(methyl methacrylate) box and
flushed with high purity nitrogen gas to reduce the background
from the environmental radon concentration. The energy resolution
of the detector was estimated by using intensive
background $\gamma$ peaks with energies 238.6 keV ($^{212}$Pb),
338.7 keV ($^{228}$Ac), 463.0 keV ($^{228}$Ac), 583.2 keV
($^{208}$Tl), 661.7 ($^{137}$Cs), 727.3 keV ($^{212}$Bi), 911.2
keV ($^{228}$Ac), 1460.8 keV ($^{40}$K) and 2614.5 keV
($^{208}$Tl) in the data measured with the cerium oxide sample in
the experiment \cite{Belli:2014}. It depends on the energy
$E_{\gamma}$ of the $\gamma$ quanta as
FWHM(keV)~$=\sqrt{1.41+0.00197\times E_{\gamma}}$, where
$E_{\gamma}$ is in keV. A sample of the purified Er$_2$O$_3$ with
mass 326 g, enclosed in a cylindric polystyrene box, was placed on
the HP Ge detector end cap. The sample contained
$1.43\times10^{21}$ and $1.532\times10^{23}$ nuclei of $^{162}$Er
and $^{170}$Er, respectively. The data with the sample were
accumulated over 1934 h, while the background spectrum was taken
over 1046 h. The two spectra, normalized for their time of
measurements, are presented in Fig. \ref{fig:bg}.

Some excess (in comparison to the background data) of $^{137}$Cs, $^{176}$Lu and $^{214}$Bi (daughter of
$^{226}$Ra) was
observed; this allowed the estimate of the residual contamination of
the sample by these radionuclides. The
activities of the nuclides in the Er$_2$O$_3$ sample after the
purification are presented in Table \ref{tab:rcont}.
The contamination by $^{176}$Lu remained almost the same as before the
purification due to the high chemical affinity between Er and Lu,
while the activity of $^{226}$Ra decreased by a factor 5.

\subsection{Search for $2\varepsilon$ and $\varepsilon\beta^{+}$ processes in $^{162}$Er}

There are no peculiarities in the energy spectrum accumulated with
the Er$_2$O$_3$ sample that could be identified as double beta
decay of the erbium isotopes. Therefore, the data were analyzed to
estimate half-life limits for the $2\varepsilon$ and
$\varepsilon\beta^+$ decay of $^{162}$Er, and the $2\beta$ decay
of $^{170}$Er. Lower half-life limits were estimated with the help
of the following equation:

\begin{center}
$\lim T_{1/2} = N \cdot \eta \cdot t \cdot \ln 2 / \lim S,$
\end{center}

\noindent where $N$ is the number of nuclei of interest in the
sample, $\eta$ is the detection efficiency (the yields of the
$\gamma$ quanta expected in the double beta processes are
included), $t$ is the time of measurement, and $\lim S$ is the
upper limit on the number of events of the effect searched for
that can be excluded at a given confidence level (C.L.). In the
present work all the $\lim S$ and, therefore, the half-life limits
are estimated at 90\% C.L. The detection efficiencies to the
effects searched for were simulated by Monte Carlo code using
EGSnrc \cite{EGSnrc} package\footnote{It should be stressed that
the calculations of the detection efficiencies with the help of
the GEANT4 simulation package give similar results
with a deviation of less than 16\% in the worst
case of the $2\nu 2K$ decay of $^{162}$Er}.
with initial kinematics given by the DECAY0 event generator
\cite{DECAY0a,DECAY0b}.

\begin{figure}[htb]
 \begin{center}
  \mbox{\epsfig{figure=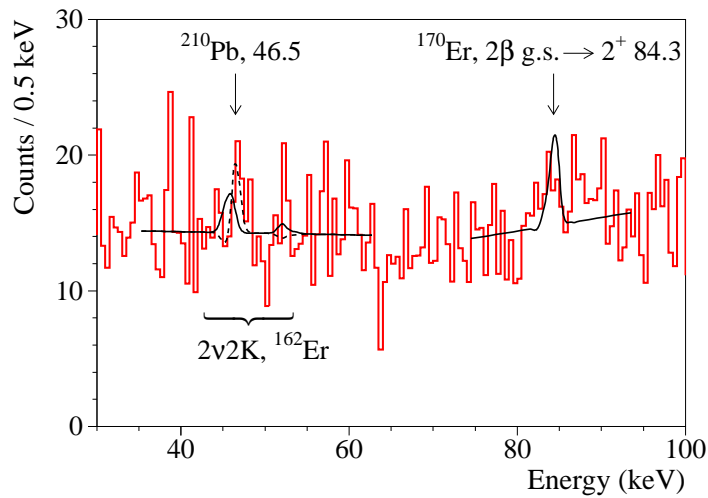,height=6.5cm}}
\vspace{-0.3cm} \caption{(Color online) Low energy part of the
spectrum accumulated with the Er$_2$O$_3$ sample over 1934 h.
The approximation function (dashed line) and the excluded effect of
$2\nu2K$ decay of $^{162}$Er (solid line) are shown. The excluded
peak of $2\beta$ decay of $^{170}$Er to the excited $2^{+}$ level
of $^{170}$Yb with the energy 84.3 keV is also shown. The energies
of the peaks are in keV.}
 \label{fig:2n2K}
 \end{center}
 \end{figure}

\begin{figure}[htb]
 \begin{center}
  \mbox{\epsfig{figure=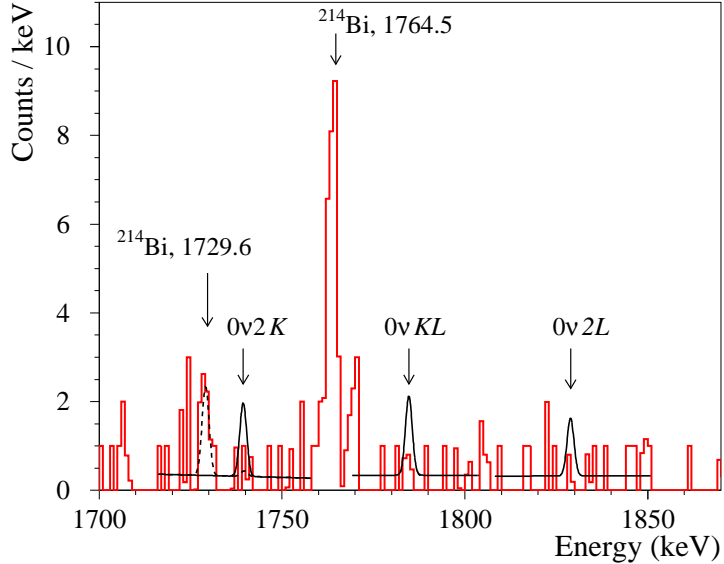,height=7.5cm}}
\vspace{-0.3cm} \caption{(Color online) Part of the energy
spectrum accumulated with the Er$_2$O$_3$ sample over 1934 h, where
the $\gamma$ peaks from the $0\nu 2K$, $0\nu KL$, and $0\nu 2L$
captures in $^{162}$Er to the ground state of $^{162}$Dy are
expected. The excluded peaks at 90\% C.L. are shown by solid
lines. The fit in the energy interval $(1716-1758)$ keV, that
includes also the $\gamma$ peak of $^{214}$Bi with energy 1729.6 keV,
is shown by a dashed line. The energies of the peaks are in keV.}
 \label{fig:0n2e}
 \end{center}
 \end{figure}

In the case of the $2\nu2K$ capture in $^{162}$Er, a cascade of X
rays and Auger electrons with energies up to 53.8 keV is expected.
We took into account only the most intense X rays of dysprosium
\cite{ToI98}: 45.2 keV (the yield of the X rays quanta is 26.8\%),
46.0 keV (47.5\%), 51.9 keV (4.9\%), 52.1 keV (9.6\%), and 53.5
keV (3.2\%). The energy spectrum accumulated with the Er$_2$O$_3$
sample was fitted by the sum of five Gaussian functions ($2\nu2K$
decay of $^{162}$Er), a peak of $^{210}$Pb with energy 46.5 keV,
and a straight line to describe the continuous background. The
best fit was achieved in the energy interval ($35-63$) keV with
$\chi^2/$n.d.f.$\simeq 0.59$, where n.d.f. is number of degrees of
freedom. The fit provides the area of the $2\nu2K$ effect: 
($-6\pm10$) counts. Taking into account the recommendations given
in \cite{Feldman:1998}, we took 11 counts as $\lim S$. The energy
spectrum in the vicinity of the $2\nu2K$ effect, the approximation
by the model of background and the excluded effect are presented
in Fig. \ref{fig:2n2K}. 
%
In this case  the detection efficiency of the whole
effect is calculated as: $\eta = \Sigma_i \ \eta_i$,
where $\eta_i$ are the efficiencies for the X ray quanta. The
detection efficiency was simulated by Monte Carlo code as $\eta =
0.016\%$. Taking into account the number of $^{162}$Er nuclei in
the sample, one can obtain the half-life limit on the $2\nu2K$
capture in $^{162}$Er presented in Table \ref{table:2blimits}.
\footnote{It should be stressed
that a possible effect of systematic errors on the obtained limit
is rather weak. E.g., the error of the efficiency calculations
(16\%, the highest one in the case of the two neutrino double $K$
capture, estimated from the difference between the simulations by
using the EGSnrc and GEANT4 codes), is negligible in comparison to
the statistical fluctuations of the excluded peak and its sigma.
The above consideration is even more valid for all other limits
reported below since the difference between the EGSnrc and GEANT4
detection efficiencies is smaller for all other double beta decay
modes and channels analyzed in the present study. The contribution
of other possible systematic errors, e.g., of the energy
calibration and resolution uncertainties, are even smaller.}
In the $0\nu$ double electron capture in $^{162}$Er (we consider
here only capture from $K$ and $L$ shells) to the ground state of
$^{162}$Dy, we assume the energies of the $\gamma$ quanta to be
equal to $E_{\gamma}=Q_{2\beta}-E_{b1}-E_{b2}$, where $E_{bi}$ are
the binding energies of the captured electrons on the atomic
shells of the daughter dysprosium atom. The energy spectrum
accumulated with the Er$_2$O$_3$ sample was fitted by the sum of a
Gaussian function (to describe the peak expected) and of a polynomial
function of the first degree (to describe the background). 
The spectrum in the vicinity of the expected peaks is shown in Fig.
\ref{fig:0n2e}. In the case of the $0\nu2K$ decay of $^{162}$Er,
also the peak of $^{214}$Bi with energy 1729.6 keV was included in
the fit to approximate the background in a wide enough energy
interval around the peak searched for. The fit
gives an area of expected $0\nu 2K$ peak with energy 1739.4 keV
$S=(0.3\pm0.9)$ counts, that corresponds to $\lim S=1.8$ counts
according to the recommendations \cite{Feldman:1998}. However, we
have used another, a more conservative approach (also recommended
in \cite{Feldman:1998} for experimental sensitivity estimated for
expected background and no true signal. Taking into account that
there are 2 counts in the energy interval of the expected peak
with energy 1739.4 keV, one should accept $\lim S=3.9$ counts (see
Table XII in \cite{Feldman:1998}). To estimate $\lim S$ for an
expected $0\nu KL$ ($0\nu 2L$) peak with energy 1784.7 keV (1828.9
keV) we have utilized the recommendations \cite{Feldman:1998} for
measured mean of a Gaussian and its sigma. The fit gives an area
of the peak $S=(1.7\pm1.5)$ ($1.1\pm1.2$) counts that corresponds to
$\lim S=4.2$ ($\lim S=3.1$) counts. The excluded peaks of the
$0\nu 2K$, $0\nu KL$, and $0\nu 2L$ captures in $^{162}$Er to the
ground state of $^{162}$Dy are shown in Fig. \ref{fig:0n2e}. The
obtained half-life limits are given in Table \ref{table:2blimits}.

 \clearpage
\begin{table*}[ht]
\vspace{-0.7cm} \caption{Half-life limits on 2$\beta$ processes in
$^{162}$Er and $^{170}$Er.}
\begin{center}
\resizebox{1.20\textwidth}{!}{
\begin{tabular}{|l|l|l|l|l|l|l|}
\hline
 Process                    & Decay     & Level of      & $E_\gamma$  & Detection   & $\lim S$  & Experimental \\
 of decay                   & mode      & daughter      & (keV)       & efficiency  & ~         & limit (yr) \\
 ~                          & ~         & nucleus       &             & (\%)        &  ~        & at 90\% C.L. \\
 ~                          & ~         & (keV)         &             & ~           &  ~        &  ~\\
 \hline
 $^{162}$Er$~\to$$^{162}$Dy & ~         & ~             & ~           & ~           &  ~        &  ~\\

 $2K$                       & $2\nu$    & g.s.          & $45-53$     & 0.016       & 11        & $\geq3.2\times10^{15}$ \\
 $2\varepsilon$             & $2\nu$    & $2^{+}~80.7$  & 80.7        & 0.014       & 2.6       & $\geq1.2\times10^{16}$ \\
 $2\varepsilon$             & $2\nu$    & $2^{+}~888.2$ & 888.2       & 1.25        & 6.5       & $\geq4.2\times10^{17}$ \\
 $2\varepsilon$             & $2\nu$    & $0^{+}~1400.3$& 1319.6      & 2.03        & 3.3       & $\geq1.3\times10^{18}$ \\
 $2\varepsilon$             & $2\nu$    & $2^{+}~1453.5$& 1187.8      & 0.86        & 6.0       & $\geq3.1\times10^{17}$ \\
 $2\varepsilon$             & $2\nu$    & $0^{+}~1666.3$& 1585.6      & 1.96        & 5.6       & $\geq7.7\times10^{17}$ \\
 $2\varepsilon$             & $2\nu$    & $2^{+}~1728.3$& 1647.6      & 0.99        & 2.3       & $\geq9.4\times10^{17}$ \\
 $KL$                       & $2\nu$    & $2^{+}~1782.7$& 1702.1      & 0.53        & 2.3       & $\geq5.0\times10^{17}$ \\
 ~                         & ~           & ~             & ~           & ~          & ~         & ~                     \\
 $2K$                       & $0\nu$    & g.s.          & $1739.1-1739.7$ & 1.87    & 3.9       & $\geq1.0\times10^{18}$ \\
 $KL$                       & $0\nu$    & g.s.          & $1783.8-1785.7$ & 1.84    & 4.2       & $\geq9.6\times10^{17}$ \\
 $2L$                       & $0\nu$    & g.s.          & $1828.6-1831.7$ & 1.82    & 3.1       & $\geq1.3\times10^{18}$ \\
 $2K$                       & $0\nu$    & $2^+$  80.7   & 1658.7        & 1.93      & 6.8       & $\geq6.2\times10^{17}$ \\
 $2K$                       & $0\nu$    & $2^{+}~888.2$ & 851.2         & 2.38      & 8.8       & $\geq5.9\times10^{17}$ \\
 $2K$                       & $0\nu$    & $0^{+}~1400.3$& 339.1         & 3.04      & 5.1       & $\geq1.3\times10^{18}$ \\
 $2K$                       & $0\nu$    & $2^{+}~1453.5$& 285.9         & 2.86      & 6.9       & $\geq9.1\times10^{17}$ \\
 $2K$                       & $0\nu$    & $0^{+}~1666.3$& 1585.6        & 1.98      & 5.6       & $\geq7.7\times10^{17}$ \\
 $2K$                       & $0\nu$    & $2^{+}~1728.3$& 1647.6        & 0.98      & 2.3       & $\geq9.3\times10^{17}$ \\
 Resonant $KL_1$            & $0\nu$    & $2^{+}~1782.7$& 1702.1        & 0.53      & 2.3       & $\geq5.0\times10^{17}$ \\
 $\varepsilon\beta^+$       & $2\nu$    & g.s.          & 511           & 6.48      & 37        & $\geq3.8\times10^{17}$ \\
 $\varepsilon\beta^+$       & $2\nu$    & $2^{+}~80.7$  & 511           & 6.48      & 37        & $\geq3.8\times10^{17}$ \\
 $\varepsilon\beta^+$       & $0\nu$    & g.s.          & 511           & 6.29      & 37        & $\geq3.7\times10^{17}$ \\
 $\varepsilon\beta^+$       & $0\nu$    & $2^{+}~80.7$  & 511           & 6.29      & 37        & $\geq3.7\times10^{17}$ \\
 ~                          & ~         & ~             & ~             & ~         & ~         & ~                     \\
 $^{170}$Er$~\to^{170}$Yb   & ~         & ~             & ~             & ~         & ~         & ~                     \\
 $2\beta^{-}$               &$2\nu+0\nu$& $2^+$ 84.3    & 84.3          & 0.017     & 9.6       & $\geq4.1\times10^{17}$  \\

 \hline
\end{tabular}
 }
\end{center}
\label{table:2blimits}
\end{table*}

\clearpage

The double electron capture in $^{162}$Er is also allowed to
excited levels of $^{162}$Dy with subsequent emission of gamma
quanta that can be detected by the HP Ge spectrometer. In the
$2\varepsilon$ process, the $2\nu$ and $0\nu$ modes cannot be
distinguished\footnote{In the present study we consider only
$0\nu2K$ transitions to the excited levels expected to be the
dominant channels of the decays.}. However, the detection
efficiencies for the decays are slightly different. The difference
is due to emission of additional $\gamma$ quanta in the $0\nu$
process with energy $E_\gamma = Q_{2\beta} - 2E_{K} - E_{exc}$,
where $E_{exc}$ is energy of the excited level of $^{162}$Dy,
and $E_{K}$ is the binding energy of the captured electrons on the $K$
atomic shell of the daughter dysprosium atom. The emission of the
$\gamma$ quanta will result in a small difference in the obtained
half-life limits\footnote{In some cases the $0\nu2K$ limits are
substantially stronger due to expected intense $\gamma$ quanta
with energy $E_{\gamma}=Q_{2\beta}-2E_{K} - E_{exc}$.}. To
estimate limits on the $2\nu2\varepsilon$ and $0\nu2K$ decays of
$^{162}$Er to the $0^{+}$ and $2^{+}$ excited levels of $^{162}$Dy
(see Fig. \ref{fig:162er}), the energy spectrum accumulated with
the Er$_2$O$_3$ sample was fitted in the energy intervals where
intense $\gamma$ peaks from the de-excitation process are
expected. The obtained limits for the double electron capture of
$^{162}$Er to the excited levels of $^{162}$Dy are presented in
Table \ref{table:2blimits}.

The $0\nu2\varepsilon$ capture in $^{162}$Er to the $2^{+}$
excited level of $^{162}$Dy with the energy $E_{\gamma}=1782.7$
keV could be much faster due to a resonant enhancement of the
capture rate. However, the recent high precise measurements of the
$^{162}$Er $Q_{2\beta}$ value by the Penning-trap mass-ratio
method have shown that the difference $Q_{2\beta} - E_K - E_L -
E_{exc} = 2.7$ keV is too big to result in a substantial resonant
enhancement of the decay probability \cite{Eliseev:2011a}.
Nevertheless, we have estimated a limit on the $0\nu2\varepsilon$
decay of $^{162}$Er to the $2^{+}$ 1782.7 keV excited level of
$^{162}$Dy as $T_{1/2}\geq5.0\times10^{17}$ yr.

\begin{figure}[htb]
\begin{center}
 \mbox{\epsfig{figure=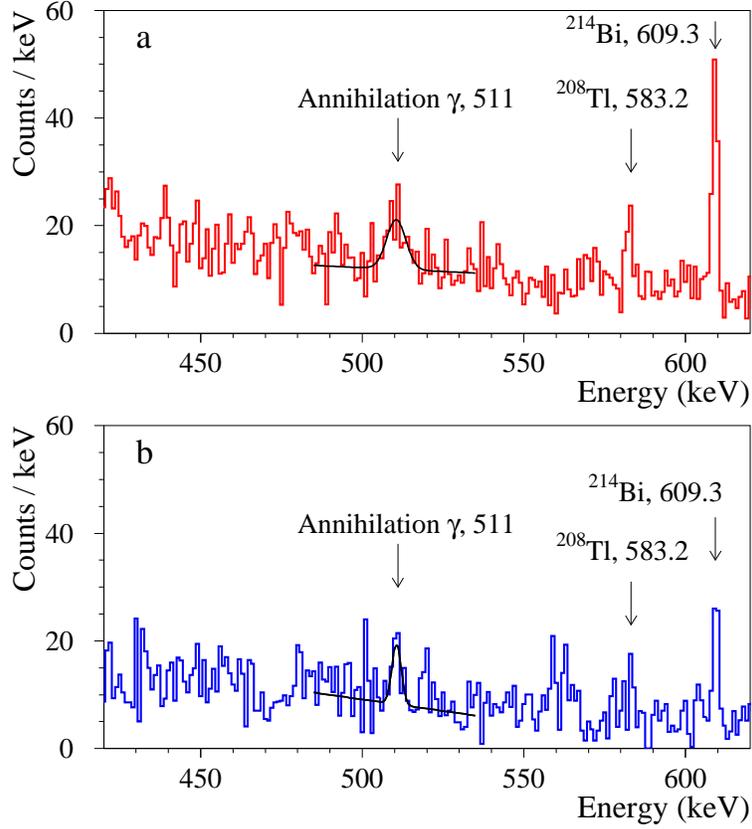,height=11.0cm}}
\caption{Energy spectra measured with the erbium oxide sample 
over 1934 h (a) and background over 1046 h (b) in the vicinity of the
511 keV annihilation peak.}
 \label{fig:511}
 \end{center}
 \end{figure}

One positron can be emitted in the $2\nu\varepsilon\beta^{+}$
($0\nu\varepsilon\beta^{+}$) decay of $^{162}$Er with an energy up
to $\approx$825 keV (depending on the binding energy
of the atomic shell of the daughter atom). The annihilation of the positron should
produce two 511 keV $\gamma$ quanta resulting in an extra counting
rate in the annihilation peak. A similar signature (annihilation
$\gamma$ quanta with energy 511 keV) is expected also in the case of
$\varepsilon\beta^{+}$ decay of $^{162}$Er to the first
$2^{+}~80.7$ keV excited level of $^{162}$Dy. To estimate $\lim S$
for the decay, the energy spectra accumulated with the Er$_2$O$_3$
sample and the background data were fitted in the energy interval
($485-535$)~keV (see Fig. \ref{fig:511}). There are ($-6\pm26$)
events in the 511 keV peak in the data accumulated with the erbium
oxide sample (taking into account the area of the annihilation
peak in the background). Since there is no evidence of the effect searched for, we
took $\lim S=37$ counts and set the limits $T_{1/2}^{\varepsilon
\beta^+}\geq 3.8\times10^{17}$ yr ($3.7\times10^{17}$ yr) on two
neutrino (neutrinoless) $\varepsilon\beta^{+}$ decay of $^{162}$Er
to the ground state and the first $2^{+}~80.7$ keV excited state
of $^{162}$Dy \footnote{It should
be noted that the half-life limits obtained by analysis of
possible 80.7 keV peak are substantially weaker due to the much
lower detection efficiencies on the level of $\sim0.01\%$.}.

\nopagebreak

\subsection{Search for  $2\beta^-$ decay of $^{170}$Er to the first $2^+$
84.3 keV excited level of $^{170}$Yb}

The double beta decay of $^{170}$Er is possible to the ground
state and to the first $2^+$ excited level of $^{170}$Yb with
energy 84.3 keV. In our experiment only the transition to the
excited level could be detected. The energy spectrum acquired with
the erbium oxide sample was fitted in the energy interval $(74-94)$
keV by a model consisting of a Gaussian function centered at 84.3
keV (to describe the effect searched for) and a straight line as a
background model. The fit (see Fig. \ref{fig:2n2K}) gives an area
of the 84.3 keV peak of $(4.7\pm 3.0)$ counts; there is no evidence
for the effect searched for. Therefore, according to
\cite{Feldman:1998} we took $\lim S=9.6$ counts. Taking into
account the number of $^{170}$Er nuclei in the sample, and the
detection efficiency $\eta=0.017\%$, we have set the limit on the
$2\beta$ decay of $^{170}$Er to the first $2^+$ excited level of
$^{170}$Yb: $T_{1/2}^{2\beta^-}\geq 4.1\times10^{17}$ yr. The
limit is for the sum of the $2\nu$ and $0\nu$ modes, since they
cannot be distinguished with the $\gamma$-spectrometry method. The
limit is slightly stronger than the one ($T_{1/2}^{2\beta^-}\geq
3.2\times 10^{17}$ yr) obtained in the experiment
\cite{Derbin:1996} with a similar technique\footnote{It should be
noted, that the result in \cite{Derbin:1996} is given with 68\%
C.L. while the present limit is estimated at 90\% C.L.}.

\section{CONCLUSIONS}

The double electron capture and the electron capture with positron
emission in $^{162}$Er, and the double beta decay of $^{170}$Er to the
first $2^+$ excited level of $^{170}$Yb were searched for in a
highly purified 326 g Er$_2$O$_3$ sample using ultra-low
background HP Ge $\gamma$ spectrometer with volume of 465 cm$^3$
at the STELLA facility of the Gran Sasso underground laboratory.
For the first time, limits on different modes and channels of
double beta decay of $^{162}$Er were set at the level of
$T_{1/2}>10^{15}-10^{18}$ yr. A possible resonant neutrinoless
double-electron capture in $^{162}$Er to the $2^{+}$ 1782.7 keV
excited level of $^{162}$Dy was restricted at the level of
$T_{1/2}\geq 5.0\times10^{17}$ yr. The sensitivity
is a few orders of magnitude weaker in comparison to the most
sensitive ``double beta plus'' experiments that already reached a
level of $\lim T_{1/2}\sim 10^{21}-10^{22}$ yr with $^{36}$Ar
\cite{Agostini:2016}, $^{40}$Ca \cite{Angloher:2016}, $^{58}$Ni
\cite{Lehnert:2016a}, $^{64}$Zn \cite{Belli:2011}, $^{78}$Kr
\cite{Gavrilyuk:2013}, $^{96}$Ru \cite{Belli:2013b}, $^{106}$Cd
\cite{Belli:2016}, $^{112}$Sn \cite{Barabash:2011}, $^{120}$Te
\cite{Andreotti:2011}, $^{124}$Xe \cite{Abe18}, $^{126}$Xe
\cite{Abe18,Gav15c}, $^{130}$Ba \cite{Meshik:2001,Pujol:2009} and
$^{132}$Ba \cite{Meshik:2001}. A new improved half-life limit
$T_{1/2}\geq4.1\times 10^{17}$ yr was set on the $2\beta^-$ decay
($2\nu + 0\nu$ modes) of $^{170}$Er to the first $2^+$ 84.3 keV
excited state of $^{170}$Yb. A typical sensitivity
to the $2\beta$ decays to $2^+$ excited levels of daughter nuclei
is on the level of $\lim T_{1/2}\sim 10^{21}-10^{25}$ yr
\cite{Barabash:2017}. It should be stressed that the $2\nu2\beta$
decay to the first excited $0^+$ levels of daughter nuclei is
observed in $^{100}$Mo and $^{150}$Nd with $T_{1/2}\sim
10^{20}-10^{21}$ yr \cite{Barabash:2017}.

A method of erbium purification from radioactive contamination
based on the liquid-liquid extraction was developed. The obtained
purified material is quite radiopure (as for other
lanthanide elements that are typically contaminated by U and Th).
Traces of $^{176}$Lu, $^{137}$Cs and $^{226}$Ra were observed in the purified
Er$_2$O$_3$ at the  $\sim (1-4)$ mBq/kg level, while other
contaminations, in particular $^{40}$K and $^{228}$Th, are below
the measurement's sensitivity of $\sim1$ mBq/kg.

\section{ACKNOWLEDGEMENTS}

The group from the Institute for Nuclear Research (Kyiv, Ukraine)
was supported in part by the program of the National Academy of
Sciences of Ukraine ``Fundamental research on high-energy physics
and nuclear physics (international cooperation)''.

\end{document}